\documentclass{elsart}
\def\be{\begin{equation}}
\def\ee{\end{equation}}
\def\bea{\begin{eqnarray}}
\def\eea{\end{eqnarray}}

\def\g5{\gamma_5}
\def\vep{\varepsilon}
\def\ot{(1\,\leftrightarrow\,2)}

\def\Journal#1#2#3#4{{#1} {\bf #2} (#3) #4}

\def\NPA{{\em Nucl. Phys.} A}
\def\NPB{{\em Nucl. Phys.} B}
\def\PRL{\em Phys. Rev. Lett.}
\def\PRC{{\em Phys. Rev.} C}
\def\PRD{{\em Phys. Rev.} D}
\def\FBS{\em Few--Body Systems}
\def\PLB{{\em Phys. Lett.} B}
\def\EPJC{{\em Eur. Phys. J.} C}
\def\ZPA{{\em Z. Phys.} A}
\def\PR{\em Phys. Rep.}
\def\fot{\frac{1}{2}}

\begin{document}
\begin{frontmatter}
\title
{Electromagnetic isoscalar $\rho \pi \gamma$ exchange current and the
anomalous action}
\author[Prague]{E. Truhl\'{\i}k}, 
\author[Prague]{J. Smejkal} and
\author[Edmonton]{F.C. Khanna}
\address[Prague]{Institute of Nuclear Physics, Academy of Sciences of the Czech Republic, CZ-25068
$\check{R}$e$\check{z}$, Czechia
\thanksref{emailT}}
\thanks[emailT]{E-mail: truhlik@ujf.cas.cz, smejkal@ujf.cas.cz}
\address[Edmonton]{Theoretical Physics Institute, Department of Physics, University
of Alberta, Edmonton, Alberta,Canada,T6G 2J1 and
TRIUMF, 4004 Wesbrook Mall, Vancouver, BC, Canada, V6T 2A3 
\thanksref{emailK}}
\thanks[emailK]{E-mail: khanna@phys.ualberta.ca}

\begin{abstract}
Using modern data, we first refit constants needed to complete 
an anomalous $\pi \rho \omega a_1$ Lagrangian 
obtained within the approach of hidden local symmetries.
Then we derive from this Lagrangian electromagnetic isoscalar $\rho \pi \gamma$
and $\rho a_1 \gamma$ exchange currents needed in calculations of the 
deuteron electromagnetic form factors at large momentum transfers. 
\end{abstract}


\end{frontmatter}

\section{Introduction}

In calculations of the electromagnetic form factors of the deuteron, the $\rho \pi \gamma$
exchange current plays an important role \cite{P&Co,MR,SMS}. According to \cite{A,CMR,GH}, it 
is derived from a $\rho \pi \gamma$ vertex
\be
<\pi^m(q_2)|J^{\,e.m.}_\lambda|\rho^l_\mu(q_1)>\,=\,i\frac{e\,g_{\rho \pi \gamma}
K_{\rho \pi \gamma}(q^2)}{m_\rho}\,\vep_{\,\lambda \nu \sigma \mu}\,q_{2\,\nu}\,
q_\sigma\,\delta_{\,ml}\,,  \label{vrpg}
\ee
where $q=q_2-q_1$ and 
the constant $g_{\rho \pi \gamma}$ is extracted  from the data on the width of
the decay $\rho\,\rightarrow\,\pi\,+\,\gamma$. 

Here we consider the task of constructing this current by using an anomalous Lagrangian.
Such a Lagrangian of the $\pi \rho \omega a_1 f_1$ system was constructed in \cite{KM}
within the approach of hidden local symmetries . It includes  also the external electromagnetic
field. The subsystem $\pi \rho \omega a_1$ was considered later in \cite{STK}, with both
the external electromagnetic and weak vector and axial--vector fields included, however. 
In the Lagrangian, several
terms are present, which yield, besides the current (\ref{vrpg}), a correction to it and also 
a new current $\rho a_1 \gamma$. The Lagrangian is characterized by several parameters, 
which should be
determined from the data on various reactions. Experimental data available 10 years ago
allowed Kaiser and Meissner \cite{KM} to extract only some of them. The
present experimental situation \cite{PDG} and results of the recent work \cite{KKW} 
allowed us to 
improve the analysis. As a result, all 4 parameters entering the 
Lagrangian of the system  $\pi \rho \omega a_1$ are now available. 

We use further the  constructed anomalous Lagrangian to derive the electromagnetic 
isoscalar $\rho \pi \gamma$ and $\rho a_1 \gamma$ exchange currents, which can be
employed in calculations of the electromagnetic form factors of the deuteron. They differ
from the standard approach by the presence of an additional momentum  dependence
in the electromagnetic form factor of the $\rho \pi \gamma$
current and by the completely new $\rho a_1 \gamma$ current. We would like to stress
that the considered model fully respects chiral invariance and vector dominance and it
is consistent with the
present experimental knowledge of elementary processes such as
radiative decays $\rho\,\rightarrow\,\pi \gamma$, $\omega\,\rightarrow\,\pi \gamma$,
$f_1\,\rightarrow\,\rho  \gamma$ and $f_1\,\rightarrow\,\rho \pi \pi$.

In Sect.\,\ref{CH1}, we introduce the anomalous Lagrangian, extract the needed constants
from the data and present the electromagnetic form factors in our model for the
vertices $\rho \pi \gamma$ and $\rho a_1 \gamma$. In Sect.\,\ref{CH2}, we construct the
associated electromagnetic isoscalar exchange currents and make the non--relativistic 
reduction of these currents.
The discussion of the results and our conclusions are given in Sect.\,\ref{reco}. 

\section{Anomalous Lagrangian of the $\pi \rho \omega a_1$ system}
\label{CH1}

We derive the $\rho \pi \gamma$ and $\rho a_1 \gamma$ vertices and the associated exchange current 
from an anomalous 
$\pi \rho \omega a_1$ Lagrangian \cite{KM,STK}. Besides the meson fields,
the external vector isoscalar ${\mathcal B}_\mu$ 
and isovector $\vec {\mathcal V}_\mu$
and  axial vector isovector $\vec {\mathcal A}_\mu$ fields are included. The Lagrangian, 
which is of interest for the problem under investigation, reads
\bea
\bar {\mathcal L}_7\,&=&\,2ig\vep_{\kappa \lambda \mu \nu}\,\{
\partial_\kappa \omega_\lambda\,[( g \vec {\rho}_\mu\,-\,e\vec {\mathcal V}_\mu)
\cdot
(\frac{1}{f_\pi}\,\partial_\nu\vec {\pi}\,+\,e\vec {\mathcal A}_\nu)] 
\nonumber  \\
& & +\,(g\omega_\kappa\,-\,\frac{1}{3}e\,{\mathcal B}_\kappa)\,
[(\partial_\lambda \vec {\rho}_\mu) \cdot (\frac{1}{f_\pi}\,
\partial_\nu \vec {\pi}\,+\,e\vec {\mathcal A}_\nu)]\}\,,  \label{LB73}  \\
\bar {\mathcal L}_8\,&=&\,-2ig\vep_{\kappa \lambda \mu \nu}\,\{
\partial_\kappa \omega_\lambda\,[( g \vec {\rho}_\mu\,-\,e\vec {\mathcal V}_\mu)
\cdot
(g \vec {a}_\nu\,+\,\frac{1}{2f_\pi} \partial_\nu \vec {\pi})] 
\nonumber  \\
& & +\,(g\omega_\kappa\,-\,\frac{1}{3}e\,{\mathcal B}_\kappa)\,
[(\partial_\lambda \vec {\rho}_\mu) \cdot (g \vec {a}_\nu\,+\,
\frac{1}{2f_\pi}\,\partial_\nu \vec {\pi})]\}\,,  \label{LB83}   \\
\bar {\mathcal L}_9\,&=&\,2ie\vep_{\kappa \lambda \mu \nu}\,\{
\frac{1}{3}\partial_\kappa {\mathcal B}_\lambda\,
[(g \vec {\rho}_\mu\,-\,e\vec {\mathcal V}_\mu) \cdot
(\frac{1}{f_\pi}\,\partial_\nu \vec {\pi}\,+\,e\vec {\mathcal A}_\nu)] \nonumber  \\
& & +\,(g\omega_\kappa\,-\,\frac{1}{3}e\,{\mathcal B}_\kappa)\,
[(\partial_\lambda \vec {\mathcal V}_\mu)(\frac{1}{f_\pi}\,\partial_\nu \vec {\pi}
\,+\,e\vec {\mathcal A}_\nu)]\}\,,  \label{LB93}   \\
\bar {{\mathcal L}}_{10}\,&=&\,-2ie\vep_{\kappa \lambda \mu \nu}\,\{
\frac{1}{3}\partial_\kappa {\mathcal B}_\lambda\,
[(g \vec {\rho}_\mu\,-\,e \vec {\mathcal V}_\mu) \cdot
(g \vec {a}_\nu\,+\,\frac{1}{2f_\pi} \partial_\nu \vec {\pi})] 
\nonumber  \\
& & +\,(g\omega_\kappa\,-\,\frac{1}{3}e\,{\mathcal B}_\kappa)\,
[(\partial_\lambda \vec {\mathcal V}_\mu) \cdot (g \vec {a}_\nu\,
+\,\frac{1}{2f_\pi} \partial_\nu \vec {\pi})]\}\,.  \label{LB103}
\eea
Here $g\,\equiv\,g_\rho$ and $e$ is the elementary charge. It follows from the KSFR relation,
$2f^2_\pi\,g^2\,=\,m^2_\rho$, that the value of g is $\approx$ 6, if $f_\pi\,=\,92$ MeV and
$m_\rho\,=\,770$ MeV are used.

The strong vertices $\rho \,\omega\,\pi$ and $\rho\,\omega\,a_1$ are already known
from the appendix A of Ref.\,\cite{KM}. 
Then Eqs.(\ref{LB73})-(\ref{LB103}) provide all other terms arising from the 
homogenous terms of the anomalous action due to the presence of
the external electroweak interactions which change the natural parity.

The total Yukawa--type anomalous Lagrangian of the $\pi \rho \omega a_1$ system is given by the sum
\be
\bar {{\mathcal L}}_{an}\,=\,\sum_{i=7,10}\,\bar{c}_i\,\bar {{\mathcal L}}_i\,, \label{Lan}
\ee
where the constants, $\bar{c}_i$, according to \cite{STK} are
\be
\bar{c}_7\,=\,\tilde{c}_7\,+\,\frac{1}{2}\tilde{c}_8\,,\quad
\bar{c}_8\,=\,\tilde{c}_8\,,\quad
\bar{c}_9\,=\,\tilde{c}_9\,+\,\frac{1}{2}\tilde{c}_{10}
\,,  \quad
\bar{c}_{10}\,=\,\tilde{c}_{10}\,, \label{bci}
\ee
and the constants $\tilde{c}_i$ are given in \cite{KM} as
\footnote[1]{The constants $\tilde{c}_i$ used here differ from those 
of Ref.~\cite{KM} by the factor g.}
\be
\hspace{-20pt}\tilde{c}_7\,=\,1.32\times 10^{-2}\,,\,\, \tilde{c}_8\,=\,-2.05\times 10^{-1}\,,\,\,
\tilde{c}_9\,=\,-5.14\times 10^{-3}\,,\,\, \tilde{c}_{10}\,=\,0\,. \label{cti}
\ee
However, new data on several reactions relevant to the analysis aiming to get these
constants has recently been published \cite{PDG} and also a new work \cite{KKW}, in addition to \cite{KM}, 
has appeared, which enables us to get more reliable values of them. But we still cannot 
remove an uncertainty in the sign of the constants, implicitly present also in the 
previous studies \cite{KM,KKW}.

\subsection{Extracting the constants from the data}

We start by extracting the constant $\tilde{c}_7$. It enters the 
effective constant $g_{\rho \omega \pi}$ defined as
\be
{\mathcal L}_{\rho \omega \pi}\,=\,i\frac{g_{\rho \omega \pi}}{f_\pi}
\vep_{\kappa \lambda \beta \nu}\partial_\kappa \omega_\lambda\vec{\rho}_\beta \cdot
\partial_\nu \vec{\pi}\,.   \label{Lrop}
\ee
The constant $g_{\rho \omega \pi}$ was  found  \cite{KKW} to have the value
\be
g_{\rho \omega \pi}\,=\,1.2\,.   \label{grompn}
\ee
On the other hand, our model Lagrangian (\ref{Lan}) leads to
\be
g_{\rho \omega \pi}\,=\,4 g^2 \tilde{c}_7\,,   \label{grompt}
\ee
which yields
\be
\tilde{c}_7\,=\,8.64 \times 10^{-3}\,.  \label{ct7n}
\ee
This value of $\tilde{c}_7$ is lower by $\sim$ 30\% than its older value
given in (\ref{cti}).

Using the Lagrangian (\ref{Lan}) we derive the amplitudes for the 
radiative processes $\rho\,\rightarrow\,\pi\gamma$ and $\omega\,\rightarrow\,\pi\gamma$ with
the constant $\tilde{c}_9$ entering in combination with $\tilde{c}_7$ into
the effective decay constants 
\be
g_{\rho \pi \gamma}\,=\,\frac{2 g m_\rho}{3 f_\pi}(\tilde{c}_7+\tilde{c}_9)\,,\quad
g_{\omega \pi \gamma}\,=\,\frac{2 g m_\omega}{ f_\pi}(\tilde{c}_7+\tilde{c}_9)\,.
\label{gro}
\ee
These decay constants enter the radiative decay width of a vector meson V decay into
a pseudoscalar meson P  as \cite{DEA}
\be
\Gamma_{V\,\rightarrow\,P \gamma}\,=\,\frac{\alpha}{24}g^{\,2}_{V P \gamma}m_V
\left[1-\left(\frac{m_P}{m_V}\right)^2 \right]^3\,.  \label{GVPg}
\ee

According to \cite{PDG}, the widths are
\be
\Gamma_{\rho^\pm\,\rightarrow\,\pi^\pm \gamma}\,=\,(67.82\,\pm\,7.54)\,keV\,,\quad
\Gamma_{\omega\,\rightarrow\,\pi^0 \gamma}\,=\,(714.9\,\pm\,42.1)\,keV\,. \label{Grocex}
\ee
From Eqs.~(\ref{ct7n})--(\ref{Grocex}) we have 
\be
\tilde{c}^c_9\,=\,8.64 \times 10^{-3}\,,\quad
\tilde{c}^n_9\,=\,9.55 \times 10^{-3}\,. \label{ct9cn}
\ee
The superscript c (n) implies that the corresponding quantity is obtained from the charged
(neutral) meson decay.
The values $\tilde{c}^c_9$ and $\tilde{c}^n_9$ differ by $\approx$ 10\%.

We calculate the weighted value of  $\tilde{c}^a_9$ according to the equation
\be
a^w\,=\,\left(\sum_{i=1,\,n}\,a_i\,w_i\right)/\left(\sum_{i=1,\,n}w_i\right)\,,
\quad w_i\,=\,\Gamma_i/\Delta\Gamma_i\,.   \label{aw}
\ee
Using Eqs.\,(\ref{Grocex}) and (\ref{ct9cn}) we obtain
\be
\tilde{c}^w_9\,=\,9.23 \times 10^{-3}\,.   \label{ct9w}
\ee

In comparison with $\tilde{c}_9$ from Eq.~(\ref{cti}), the new value (\ref{ct9w}) has
the opposite sign and it is larger by a factor of $\sim$ 2.

If we admit negative values for the sum $\tilde{c}_7\,+\,\tilde{c}_9$ then we
get
\be
\tilde{c}^w_9\,=\,-2.65 \times 10^{-2}\,. \label{ct9mw}
\ee

With the weighted value (\ref{ct9w}) we get for 
the radiative decay widths
\be
\Gamma_{\rho^\pm\,\rightarrow\,\pi^\pm \gamma}\,=\,72.5\,keV\,,\quad
\Gamma_{\omega\,\rightarrow\,\pi^0 \gamma}\,=\,689.6\,keV\,, \label{GCS}
\ee
and for their ratio
\be
\Gamma_{\omega\,\rightarrow\,\pi^0 \gamma}/\Gamma_{\rho^\pm\,\rightarrow\,\pi^\pm \gamma}
\,=\,9.5\,,  \label{Grocax}
\ee
and for the effective decay constants
\be
g_{\rho \pi \gamma}\,=\,0.585\,,\quad
g_{\omega \pi \gamma}\,=\,1.78\,.   \label{gsca}
\ee
For the negative values, the effective couplings have the opposite sign.

We can also use the same procedure with the neutral $\rho$ meson decay, for which
the decay width is \cite{PDG}
\be
\Gamma_{\rho^0\,\rightarrow\,\pi^0 \gamma}\,=\,(102.5\,\pm\,25.6)\,keV\,. \label{Gronex}
\ee
The result for the constant $\tilde{c}_9$ from this equation is
\be
\tilde{c}_9\,=\,1.25 \times 10^{-2}\,.  \label{ctn9n}
\ee
The experimental value of $\Gamma_{\rho^0\,\rightarrow\,\pi^0 \gamma}$ is converging in time to the
value of $\Gamma_{\rho^\pm\,\rightarrow\,\pi^\pm \gamma}$ \cite{PDG}, but
it is still impossible to get both decay widths at the 1$\sigma$ error level by 
using the same value of $\tilde{c}_9$, since the error bounds do not overlap.
We can use  Eq.\,(\ref{aw}) for the neutral meson
decays to get the weighted value of  $\tilde{c}^w_9\,=\,1.014 \times 10^{-2}$, but
the calculated value of
$\Gamma_{\omega\,\rightarrow\,\pi^0 \gamma}\,=\,762$ keV is already outside the 1$\sigma$
error bound 757 keV.
It can be verified that only the values of $\tilde{c}_9$ from the interval
$0.97 \times 10^{-2}\,<\,\tilde{c}_9\,<\,1.05 \times 10^{-2}$ yield simultaneously 
acceptable values of
$\Gamma_{\omega\,\rightarrow\,\pi^0 \gamma}$ and $\Gamma_{\rho^0\,\rightarrow\,\pi^0 \gamma}$.
For 
\be
\tilde{c}_9\,=\,1.00 \times 10^{-2}\,,  \label{ct9exp}
\ee
one gets 
\be
\Gamma_{\omega\,\rightarrow\,\pi^0 \gamma}\,=\,751\,keV\,,\quad
\Gamma_{\rho^0\,\rightarrow\,\pi^0 \gamma}\,=\,79.4\,keV\,,  \label{Gronex1}
\ee
with the ratio
\be
\Gamma_{\omega\,\rightarrow\,\pi^0 \gamma}/\Gamma_{\rho^0\,\rightarrow\,\pi^0 \gamma}
\,=\,9.5\,,  \label{Gronax}
\ee
and for the effective decay constants a value of
\be
g_{\rho \pi \gamma}\,=\,0.610\,,\quad
g_{\omega \pi \gamma}\,=\,1.86\,.   \label{gsna}
\ee
One obtains similar results with the negative value of $\tilde{c}_9$
\be
\tilde{c}_9\,=\,-2.73 \times 10^{-2}\,.  \label{ct9mexp}
\ee

Actually, the decay widths
$\Gamma_{\omega\,\rightarrow\,\pi^0 \gamma}$ and $\Gamma_{\rho^0\,\rightarrow\,\pi^0 \gamma}$
were used in \cite{KKW} to extract the constant d, related to our constants by the 
equation d$\,=\,g\,(\tilde{c}_7+\tilde{c}_9)$, with the result d$\,\approx\,$0.01. On the other hand,
taking our constants g, $\tilde{c}_7$  (\ref{ct7n}) and $\tilde{c}_9$ (\ref{ct9exp})
we get d$\,\approx\,$0.11. Evidently, the calculations \cite{KKW} give the value of d$^2$ instead of d.

It is interesting to follow the change of the value of the constant g$_{\omega \pi \gamma}$ in time.
The first use of the $\rho \pi \gamma$ exchange current in the calculations of the electromagnetic
deuteron form factors and an estimate of the constant
g$_{\omega \pi \gamma}$ was made by Chemtob, Moniz and Rho (CMR) in \cite{CMR}. Actually,  
CMR adopted the $\rho \pi \gamma$  exchange current constructed by Adler \cite{A}.
The existing data on 
$\Gamma_{\rho\,\rightarrow\,\pi \gamma}$ \cite{PDGVO} allowed CMR to extract the upper bound
$|$ g$_{\rho \pi \gamma}|\,<\,$1.02. Inspite of an estimate, based on the existing
relativistic quark model \cite{FKR}, that the value of   g$_{\rho \pi \gamma}$ migt be
about 50\% of the upper bound, CMR adopted the value g$_{\rho \pi \gamma}$\,=\,1.
Afterwards, the  value g$_{\rho \pi \gamma}$\,=\,0.34--0.48 was extracted  by Gari and Hyuga 
\cite{GH} from 
the more precise data \cite{G}.
Later on, Towner \cite{T} analysed the data \cite{J} to obtain the value 
g$_{\rho \pi \gamma}$\,=\,0.578, which
is larger by about 50\%.  
At last, in \cite{HT,IG}, the value of g$_{\rho \pi \gamma}$ , obtained from the data \cite{B}, 
has oscillated down to 0.56. 
However, our value (\ref{gsca}) of g$_{\rho \pi \gamma}$\,=\,0.585 is again larger and close to 
the one obtained in \cite{T}. Towner also extracted
from the data \cite{PDGO} the constant g$_{\omega \pi \gamma}$\,=\,1.98, which is larger by about 10\% 
than our g$_{\omega \pi \gamma}$\,=\,1.78 (\ref{gsca}).

Let us note that our analysis is based on the data recommended recently in \cite{PDG}
and that in our approach, we extract g$_{\rho \pi \gamma}$ and g$_{\omega \pi \gamma}$ from the data 
simultaneously. This is due to the fact that in our model, both vertices depend on the same parameters
$\tilde{c}_7$ and $\tilde{c}_9$.

We conclude that the present experimental results  allow us to extract two values of
the constant $\tilde{c}_9$ which differ by $\approx$ 10\%
\footnote[2]{In the case of the negative values, the uncertainty in $\tilde{c}_9$
is about 3\%.}. However, the difference in the effective coupling constants 
g$_{\rho \pi \gamma}$ and g$_{\omega \pi \gamma}$ given in Eqs.\,(\ref{gsca}) and (\ref{gsna})
is only 4\%.
One can also ignore the data on the  radiative decay $\rho^0\,\rightarrow\,\pi^0+\gamma$ 
and use only  $\tilde{c}_9$ from Eq.\,(\ref{ct9w}) and the decay constants
(\ref{gsca}) with the hope that the experimental study of this decay  will improve.

Let us turn to determining  the constants $\tilde{c}_8$ and $\tilde{c}_{10}$, needed to
calculate the anomalous processes with the $a_1$ meson. However, the decays of this meson
are not well measured and we use the $f_1$ meson decays to extract them from the data.
In contrast to the experimental situation in late eighties, the decay 
\mbox{$f_1\,\rightarrow\,\rho \pi \pi$} is now well measured \cite{PDG} and good data 
for the radiative decay
\mbox{$f_1\,\rightarrow\,\rho \gamma$} has appeared \cite{PDG}. According to \cite{KM}, one of the
contact couplings, $\tilde{h}'_{f_1 \rho \pi \pi}$, is connected to $\tilde{c}_8$ as
\be
\tilde{c}_8\,=\,-\frac{\tilde{h}'_{f_1 \rho \pi \pi}}{g^2}\,.  \label{ct8t}
\ee
Using Eq.~(4.20) of \cite{KM} with the new value
\be
\Gamma_{f_1\,\rightarrow\,\rho \pi \pi}\,=\,(2.808\,\pm\,0.360)\,MeV\,,  \label{Gf1rpp}
\ee
one gets 
\be
\tilde{h}'_{f_1 \rho \pi \pi}\,=\,3.775\,,  \label{ht}
\ee
and from (\ref{ct8t})
\be
\tilde{c}_8\,=\,-1.02 \times 10^{-1}\,.  \label{ct8n}
\ee
Again, the new value of $\tilde{c}_8$ is lower by a factor of $\sim$ 2.

In order to calculate the radiative $f_1\,\rightarrow\,\rho \gamma$ decay, 
we adopt the pertinent part of the anomalous Lagrangian (3.12) 
of Ref.~\cite{KM},
\bea
\Delta {\mathcal L}_{f_1}\,&=&\,\tilde{c}_8 \Delta {\mathcal L}_9\,+\,\tilde{c}_{10} \Delta 
{\mathcal L}_{12}\,,  \label{Lf1} \\
\Delta {\mathcal L}_9\,&=&-2i g^2 \,\vep_{\kappa \lambda \beta \nu} f_{1 \kappa}[
g (\partial_\lambda \vec{\rho}_\beta) \cdot \vec{\rho}_\nu\,-\,(\partial_\lambda \rho^0)
(e \widetilde {\mathcal B}_\nu)]\,,  \label{L9}  \\
{\mathcal L}_{12}\,&=&\,-2ig^2\vep_{\kappa \lambda \beta \nu} f_{1 \kappa} (\partial_\lambda
e \widetilde {\mathcal B}_\beta) \rho^0_\nu\,.  \label{L12}
\eea
With these vertices, we get for the radiative decay width an equation similar to Eq.\,(\ref{GVPg})
\be
\Gamma_{f_1\,\rightarrow\,\rho \gamma}\,=\,\frac{\alpha}{24}g^2_{\rho f_1 \gamma}\,m_{f_1}\,
\left[1\,+\,\left(\frac{m_{f_1}}{m_\rho}\right)^2\right]
\left[1\,-\,\left(\frac{m_\rho}{m_{f_1}}\right)^2\right]^3\,,  \label{Gf1rgt}
\ee
where
\be
g_{ \rho f_1 \gamma}\,=\,2 g^2\,\left( \tilde{c}_8\,+\,\tilde{c}_{10}\right)\,.  \label{gf1rg}
\ee
Using the experimental value of $\Gamma_{f_1\,\rightarrow\,\rho \gamma}$ \cite{PDG}
\be
\Gamma_{f_1\,\rightarrow\,\rho \gamma}\,=\,(1.296\,\pm\,0.288)\,MeV\,,  \label{Gf1rge}
\ee
we get
\be
\tilde{c}_8\,+\,\tilde{c}_{10}\,=\,2.65 \times 10^{-2}\,,   \label{ct810}
\ee
and for the constant $\tilde{c}_{10}$ the value
\be
\tilde{c}_{10}\,=\,1.29 \times 10^{-1}\,,  \label{ct10n}
\ee
which is in absolute value even larger than $\tilde{c}_8$, Eq.\,(\ref{ct8n}).
Then for the effective coupling $g_{f_1 \rho \gamma}$ we have the value
\be
g_{f_1 \rho \gamma}\,=\,1.84\,.  \label{gf1rgn}
\ee
Admitting negative values for the sum $\tilde{c}_8+\tilde{c}_{10}$ we get
\be
\tilde{c}_{10}\,=\,7.59 \times 10^{-2}\,.  \label{ct10mn}
\ee

We now have in Eqs.~(\ref{ct7n}),(\ref{ct9w}),(\ref{ct9exp}),(\ref{ct8n}) and 
(\ref{ct10n}) all the needed
constants and we can now proceed to constructing the $\rho \pi \gamma$ and
$\rho a_1 \gamma$ exchange currents.
However, let us first discuss the electromagnetic form factors provided by our model.

\subsection{Electromagnetic form factors}

The Lagrangian of Eq.~(\ref{Lan}) yields the vertex
$\rho \pi \gamma$, Eq.\, (\ref{vrpg}), with the form factor
\be
K_{\rho \pi \gamma}(q^2)\,=\,F_\omega(q^2)\,+\,\left(\frac{\tilde{c}_9-\tilde{c}_7}
{\tilde{c}_9+\tilde{c}_7}\right)\left[1\,-\,F_\omega(q^2)\right]\,,
  \label{rpgff}
\ee
where the form factor $F_\omega(q^2)$ is given by the vector dominance. Since 
$\tilde{c}_9\,>\,\tilde{c}_7$, the second term on the
right hand side of Eq.\,(\ref{rpgff}) gives a 3--7\% positive correction to the first term 
at $\vec{q}^{\,2}\,\approx\,m^2_\omega$.
In the case of the negative values of $\tilde{c}_9$, this term is larger by a factor 
of $\sim$ 2 than the first one at these values of $\vec{q}^{\,2}$.

The presence of the terms $\bar {\mathcal L}_8$ and $\bar {\mathcal L}_{10}$ in Eq.\,(\ref{Lan})
shows the existence of the vertices
$\rho a_1 \gamma$,
which give rise to the corresponding $\rho a_1 \gamma$
currents. In analogy with Eq.~(\ref{vrpg}) we have 
\be
\hspace{-20pt}<a_{1\,\nu}^m(q_2)|J^{\,e.m.}_\lambda|\rho^l_\mu(q_1)>\,=\,e g_{\rho a_1 \gamma}
\vep_{\,\lambda \nu \sigma \mu}\,\left[K_1(q^2)q_\sigma\,+\,K_2(q^2)q_{1\,\sigma}\right]\,
\delta_{\,ml}\,,  \label{vra1g}
\ee
where the form factors $K_{1,2}(q^2)$ are defined as
\be
\hspace{-15pt}K_1(q^2)=\frac{1}{\tilde{c}_8+\tilde{c}_{10}}\left[\tilde{c}_8 F_\omega(q^2)+\tilde{c}_{10}
\right]\,,\quad 
K_2(q^2)=\frac{\tilde{c}_8}{\tilde{c}_8+\tilde{c}_{10}}\left[F_\omega(q^2)-1\right]\,. 
 \label{K12}
\ee
The effective coupling constant $g_{\rho a_1 \gamma}$ is 
\be
g_{\rho a_1 \gamma}\,=\,g_{\rho f_1 \gamma}/3\,,  \label{gra1g}
\ee
with the constant $g_{\rho f_1 \gamma}$ given in Eq.\,(\ref{gf1rg}).
This gives a new contribution to the current.

In the next section, we derive the $\rho \pi \gamma$ and $\rho a_1 \gamma$ exchange 
currents.

\section{The $\rho \pi \gamma$ and $\rho a_1 \gamma$ exchange currents}
\label{CH2}

The general structure of the exchange currents which follows from our Lagrangian
(\ref{Lan}) is given in Fig.~1. 
The currents of the pion range are,
\bea
J^{\,m.~c.}_{\,\pi\,\mu}\,&=&\,ig_{\pi NN}\,\frac{2g^2\tilde{c}_7}{3f_\pi}\,\vep_{\mu \kappa \beta \nu}\,
F_\omega(q^2)q_\kappa {\mathcal P}_\beta(1,2)q_{2\,\nu}\,\,+\,\ot\,,  \label{Jpimc} \\
J^{\,c}_{\,\pi\,\mu}\,&=&\,-ig_{\pi NN}\,\frac{g^2(\tilde{c}_7-\tilde{c}_9)}{3f_\pi}\,
\vep_{\mu \kappa \beta \nu}\,
q_\kappa {\mathcal P}_\beta(1,2) q_{2\,\nu}\,\,+\,\ot\,,  \label{Jpic}
\eea
where
\be
{\mathcal P}_\beta(1,2)\,=\,\Gamma^a_{1\,,\beta}\,\Delta^\rho_F(q^{\,2}_1)\,
\Delta^\pi_F(q^{\,2}_2)\,\Gamma^a_{2\,,5}\,,    \label{Pb}
\ee
and
\be
\Gamma^a_{i\,,\beta}\,=\,\bar{u}(p'_i)(\gamma_\beta - \frac{\kappa_V}{2M}\sigma_{\beta 
\delta}\,q_{i\,\delta})\tau^a u(p_i)\,\,and\,\, \Gamma^a_{j\,,5}\,=\,\bar{u}(p'_j)\gamma_5
\tau^a u(p_j)\,.  \label{Gver}
\ee

The sum of the currents given  by Eqs.\,(\ref{Jpimc}) and (\ref{Jpic}) is
\be
J_{\,\pi\,\mu}\,=\,ig_{\pi NN}\frac{g}{2}\frac{g_{\rho \pi \gamma}}{m_\rho}\,K_{\rho \pi \gamma}(q^2)
\,\vep_{\mu \kappa \beta \nu}\,
q_\kappa {\mathcal P}_\beta(1,2)q_{2\,\nu}\,\,+\,\ot\,,  \label{Jpi}
\ee
where the effective coupling constant $g_{\rho \pi \gamma}$ 
and the form factor $K_{\rho \pi \gamma}(q^2)$ are defined in Eqs.\,(\ref{gro}) and (\ref{rpgff}),
respectively.

\input{feynman}
\vspace{20pt}
\hspace{70pt}
\begin{picture}(10000,15000)
\drawline\fermion[\N\REG](0,0)[13948]
\drawarrow[\N\ATBASE](0,2000)
\drawarrow[\N\ATBASE](0,11948)
\global\advance\fermionbackx by -1500
\put(\fermionbackx,\fermionbacky){$p_{\,1}^{\,\prime}$}
\put(\fermionbackx,0){$p_{\,1}$}
\put(4200,-1800){a}
\global\seglength=1400
\global\gaplength=350
\thicklines
\drawline\scalar[\E\REG](0,9948)[5]
\thinlines
\drawarrow[\E\ATBASE](6000,\pmidy)
\drawarrow[\W\ATBASE](2400,\pmidy)
\put(1800,10450){$B_1$}
\put(6300,10450){$B_2$}
\put(1800,9000){$q_{\,1}$}
\put(6300,9000){$q_{\,2}$}
\thicklines
\drawline\fermion[\S\REG](\pmidx,\pmidy)[2100]
\drawarrow[\N\ATBASE](\pmidx,\pmidy)
\global\advance\pmidx by 300
\global\advance\pmidy by -400
\put(\pmidx,\pmidy){$B$}
\thinlines
\drawline\photon[\S\REG](\fermionbackx,\fermionbacky)[5]
\global\advance\pmidx by -1600
\global\advance\pmidy by -3700
\put(\pmidx,\pmidy){${J}^{\,m.c.}_{\,B_2\,\mu}(q)$}
\drawline\fermion[\N\REG](8400,0)[13948]
\drawarrow[\N\ATBASE](8400,2000)
\drawarrow[\N\ATBASE](8400,11948)
\global\advance\fermionbackx by 600
\put(\fermionbackx,\fermionbacky){$p_{\,2}^{\,\prime}$}
\put(\fermionbackx,0){$p_{\,2}$}
\end{picture}
\hspace{50pt}
\begin{picture}(10000,15000)
\drawline\fermion[\N\REG](0,0)[13948]
\drawarrow[\N\ATBASE](0,2000)
\drawarrow[\N\ATBASE](0,11948)
\put(4200,-1800){b}
\global\advance\fermionbackx by -1500
\put(\fermionbackx,\fermionbacky){$p_{\,1}^{\,\prime}$}
\put(\fermionbackx,0){$p_{\,1}$}
\thicklines
\global\seglength=1400
\global\gaplength=350
\drawline\scalar[\E\REG](0,9948)[5]
\thinlines
\drawarrow[\E\ATBASE](6000,\pmidy)
\drawarrow[\W\ATBASE](2400,\pmidy)
\put(1800,10450){$B_1$}
\put(6300,10450){$B_2$}
\put(1800,9000){$q_{\,1}$}
\put(6300,9000){$q_{\,2}$}
\drawline\photon[\S\REG](\pmidx,\pmidy)[7]
\global\advance\pmidx by 200
\global\advance\pmidy by -1100
\drawarrow[\N\ATBASE](\pmidx,\pmidy)
\global\advance\pmidx by -1600
\global\advance\pmidy by -3700
\put(\pmidx,\pmidy){${J}^{\,c}_{\,B_2\,\mu}(q)$}
\drawline\fermion[\N\REG](8400,0)[13948]
\drawarrow[\N\ATBASE](8400,2000)
\drawarrow[\N\ATBASE](8400,11948)
\global\advance\fermionbackx by 600
\put(\fermionbackx,\fermionbacky){$p_{\,2}^{\,\prime}$}
\put(\fermionbackx,0){$p_{\,2}$}
\end{picture}
\vspace{10mm}
\newline
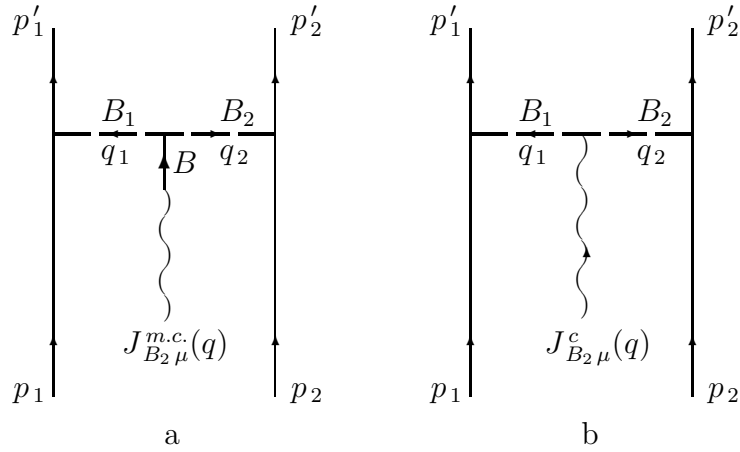
Fig.\,1. The general structure of the electromagnetic
isoscalar $\rho \pi \gamma$ and $\rho a_1 \gamma$ exchange currents $J_\mu(q)$
considered in this paper.
The meson $B_1$ is the $\rho$ meson.
The range of the
current is given by the meson $B_2$ which is either $\pi$ or $a_1$ meson.
The graph a is for a mesonic current $ J^{\,m.~c.}_{\,B_2\,\mu}$,
where the electromagnetic isoscalar 
interaction is mediated by the meson B which is $\omega$ meson.
Contact terms are given by the graph b, $ J^{\,c}_{\,B_2\,\mu}$, where the isoscalar
electromagnetic current interacts directly
with the mesons $B_1$ and $B_2$.

The currents of the $a_1$ meson range are
\bea
J^{\,m.~c.}_{\,a_1\,\mu}\,&=&\,-\frac{2 g_A}{3}\,g^4 \tilde{c}_8\,\vep_{\mu \kappa \beta \nu}\,
F_\omega(q^2)(q_\kappa\,+\,q_{1\,\kappa})\,{\mathcal P}_{\beta \nu}(1,2)\,+\,\ot\,, \label{Ja1mc} \\
J^{\,c}_{\,a_1\,\mu}\,&=&\,\frac{g_A}{3} g^4\,\vep_{\mu \kappa \beta \nu}\,
\left(\tilde{c}_8\, q_{1\,\kappa}\,-\,\tilde{c}_{10}\, q_\kappa \right)\,
{\mathcal P}_{\beta \nu}(1,2)\,+\,\ot\,, \label{Ja1c}
\eea
where the weak axial coupling constant $g_A\,=\,1.26$, and
\be
\hspace{-20pt}{\mathcal P}_{\beta \nu}(1,2)\,=\,\Gamma^a_{1\,,\beta}\,\Delta^\rho_F(q^{\,2}_1)\,
\Delta^{a_1}_F(q^{\,2}_2)\,\Gamma^a_{2\,,5\nu}\,\,and\,\,  
\Gamma^a_{j\,,5\nu}\,=\,\bar{u}(p'_j)\gamma_\nu\gamma_5\tau^a u(p_j)\,.  \label{Pbn}
\ee

The total $\rho a_1 \gamma$ current is
\be
\hspace{-20pt}J_{\,a_1\,\mu}=-\frac{g_A}{2}g^2 g_{\rho a_1 \gamma}\,\vep_{\mu \kappa \beta \nu}
\left[K_1(q^2)q_\kappa+K_2(q^2)q_{1\kappa}\right]
{\mathcal P}_{\beta \nu}(1,2)+\ot\,, \label{Ja1}
\ee
with the effective coupling constant $g_{\rho a_1 \gamma}$ and with the form factors $K_{1,2}(q^2)$ 
given in Eqs.\,(\ref{gra1g}) and (\ref{K12}), respectively.

In deriving Eqs.(\ref{Jpimc}) and (\ref{Jpic}) and Eqs.\,(\ref{Ja1mc}) and (\ref{Ja1c}), 
we use the equation $g_\omega\,=\,3g_\rho$,
which ensures the current conservation.

\subsection{Non--relativistic reduction}

Now we proceed to the non-relativistic reduction of the currents given by Eqs.\,(\ref{Jpi})
and ({\ref{Ja1}). Having in mind the application of our results 
to the elastic electron scattering by nuclei, we neglect the energy transfer putting $q_0\,=\,0$.
This reduction yields for the
space component of the  currents 
\bea
\vec{J}_\pi\,&=&\,
C_\pi
\left<\left(\vec{q} \times \vec{q}_2\right) 
 -\,\frac{1}{4M^2}\left\{\left(\fot+\kappa_V\right)
\left(\vec{q} \times \vec{q}_2\right)\left[{\vec q}^{\,2}_1 
\,+i\,{\vec \sigma}_1 \cdot
{\vec P}_1\times{\vec q}_1\right] \right. \right. \nonumber \\
&&\left. \left. -i\,\left(1+\kappa_V\right)\left({\vec q}_2\cdot {\vec P}_2\right)
\left(\vec q \times\left({\vec \sigma}_1 \times {\vec q}_1 \right)\right) \right\}\right>
 \nonumber  \\
& &\hspace{3mm}\cdot\,\left(\vec{\sigma}_2\cdot \vec{q}_2\right)\, 
(\vec{\tau}_1 
\cdot \vec{\tau}_2)\,\Delta^\rho_F(\vec{q}^{\,2}_1)\,\Delta^\pi_F(\vec{q}^{\,2}_2)\,+\,
\ot\,, \label{sjpi}  \\
\vec{J}_{a_1}\,&=&\,i\frac{g_A}{2}\,g^2 \,g_{\rho a_1 \gamma}\,\left(K_1(\vec{q}^{\,2})
\vec{q}\,+\, K_2(\vec{q}^{\,2})\vec{q}_1\right) \times \left< \vec{\sigma}_2 
\left\{1-\frac{1+2\kappa_V}{8M^2}
\left[{\vec q}^{\,2}_1 \right.\right.\right. \nonumber \\
&& \left.\left.\left.+i\,{\vec \sigma}_1 \cdot
{\vec P}_1\times{\vec q}_1\right] \right\}
-i\,\left({\vec \sigma}_1\times{\vec q}_1\right)\frac{1+\kappa_V}{4M^2}
\left({\vec \sigma}_2 \cdot {\vec P}_2 \right)\right> \nonumber \\
& & \hspace{3mm} \cdot\, (\vec{\tau}_1 
\cdot \vec{\tau}_2)\,\Delta^\rho_F(\vec{q}^{\,2}_1)\Delta^{a_1}_F(\vec{q}^{\,2}_2)\,+\,
\ot\,.   \label{sja1}
\eea
Here
\be
C_\pi\,=\,-\,i\frac{g_{\pi NN}}{2M}\,\frac{g}{2m_\rho}\,g_{\rho \pi \gamma}\,
K_{\rho \pi \gamma}(\vec q^{\,2})\,,  \label{Cpi}
\ee
and
\be
\vec{P}_i\,=\,\vec{p}'_i\,+\,\vec{p}_i\,.  \label{Pi}
\ee
The importance of relativistic corrections in the space component of the $\rho \pi \gamma$ 
current was stressed in \cite{HT}. 
They were obtained earlier in \cite{Tal}.
We present them also in the $a_1$ exchange term.
An evaluation of the contribution of these terms to the electromagnetic deuteron
form factors would give more confidence in
the validity of the static approximation for the current (\ref{sja1}).

Besides the relativistic corrections present in the current (\ref{sjpi}), we list
the boost current $\vec{J}^{\,B}_\pi$ arising from the commutator of the leading
terms in Eq.\,(\ref{sjpi}) with the operator of the kinematical boost 
and the retardation current $\vec{J}^{\,ret}_\pi$.
The kinematical boost current is defined as \cite{AA}
\be
\vec{J}^{\,B}_\pi\,=\,i\left[\chi_0\left(\vec{P}\right)\,,\,
\vec{J}^{\,0}_\pi\right]\,,  \label{sjpiB}
\ee
where
\bea
\chi_0(\vec{P})\,&=&\,-\frac{i}{16 M^2}\left[ -\vec{P}^{\,2}-2i\left(\vec{r}\cdot \vec{P}\right)
\left(\vec{p}\cdot \vec{P}\right)+2i\left(\vec{\sigma}_1-\vec{\sigma}_2\right)\times \vec{p}
\cdot \vec{P}\right]\,,  \label{kbo} \\
\vec{r}\,&=&\,\vec{r}_1-\vec{r}_2\,,  \label{rp}
\eea
and the static approximation to the $\rho \pi \gamma$ current $\vec{J}^{\,0}_\pi$ is
defined as
\be
\vec{J}^{\,0}_\pi\,=\,
C_\pi \left(\vec{q} \times \vec{q}_2\right) 
(\vec{\sigma}_2\cdot \vec{q}_2)\, 
(\vec{\tau}_1 \cdot \vec{\tau}_2)\,\Delta^\rho_F(\vec{q}^{\,2}_1)
\,\Delta^\pi_F(\vec{q}^{\,2}_2)\,+\,
\ot\,. \label{sj0pi} 
\ee
Generally, the current $\vec{J}^{\,B}_\pi$ is quite complicated. However, it simplifies
in the Breit system, where it reads in 
the c.m. coordinates,
\bea
\vec{J}^{\,B}_\pi\,&=&\,\frac{i}{16M^2}C_\pi\left(\vec{q} \times \vec{q}_- \right) 
\left\{
\left(\vec{\sigma}_2\cdot \vec{q}_- \right)
\left[-\frac{1}{4}\left(\vec{r}\cdot
\vec{q}\right)\left(\vec{q}\cdot\vec{k}\right)+\left(\vec{\sigma}_1\cdot
\vec{K}\times\vec{q}\right)\right]  \right. \nonumber \\
&&\left.\,-\left(\vec{q}_-\cdot\vec{K}\times\vec{q}\right)
+i\left(\vec{\sigma}_2\times\vec{q}_-\right)\cdot\left(\vec{k}\times\vec{q}\right)
\right\} \nonumber \\
&& \cdot \left(\vec{\tau}_1 
\cdot \vec{\tau}_2\right)\,\Delta^\rho_F\left(\vec{q}^{\,2}_+\right)
\,\Delta^\pi_F\left(\vec{q}^{\,2}_-\right)\,+\,\ot\,.    \label{sjpiBf}
\eea
Here
\bea
\vec{p}\,&=&\,\fot\left(\vec{p}_1-\vec{p}_2\right)\,,\quad
\vec{p}'\,=\,\fot\left(\vec{p}'_1-\vec{p}'_2\right)\,,\quad
\vec{P}\,=\,\vec{p}_1+\vec{p}_2\,,\quad
\vec{P}'\,=\,\vec{p}'_1-\vec{p}'_2\,,   \nonumber  \\
\vec{k}\,&=&\,\vec{p}'-\vec{p}\,,\quad 
\vec{K}\,=\,\vec{p}'+\vec{p}\,,\quad 
\vec{q}\,=\,\vec{P}'-\vec{P}\,,\quad
\vec{Q}\,=\,\vec{P}'+\vec{P}\,,   \nonumber \\
\vec{q}_1\,&=&\,\vec{p}'_1-\vec{p}_1\,=\,\vec{k}+\fot\vec{q}\,\equiv\,\vec{q}_+\,,\quad
\vec{q}_2\,=\,\vec{p}'_2-\vec{p}_2\,=\,-\vec{k}+\fot\vec{q}\,\equiv\,-\vec{q}_-\,,
\nonumber \\
\vec{P}_1\,&=&\,\vec{p}'_1+\vec{p}_1\,=\,\vec{K}+\fot\vec{Q}\,,\quad
\vec{P}_2\,=\,\vec{p}'_2+\vec{p}_2\,=\,-\vec{K}+\fot\vec{Q}\,.  \label{cmc}
\eea
In the Breit system, $\vec{Q}=\vec{P}'-\vec{P}=0$ and $\vec{P}'=-\vec{P}=\vec{q}/2$.

In the realistic calculations, the currents, Eq.\,(\ref{Jpi}) and Eq.\,(\ref{Ja1}), are
multiplied by the strong form factors. In essence, this procedure reduces in the case
of the $\rho \pi \gamma$ current to 
the following change in Eq.\,(\ref{Pb}),
\be
\Delta^\rho_F\left(\vec{q}^{\,2}_1\right)\,\Delta^\pi_F\left(\vec{q}^{\,2}_2\right)\,
\rightarrow\,F_{\rho NN}\left(\vec{q}^{\,2}_1\right)
\Delta^\rho_F\left(\vec{q}^{\,2}_1\right)\,
F_{\pi NN}\left(\vec{q}^{\,2}_2\right)
\Delta^\pi_F\left(\vec{q}^{\,2}_2\right)\,,  \label{rhopiff}
\ee
and to the change 
\be
\Delta^\rho_F\left(\vec{q}^{\,2}_1\right)\,\Delta^{\,a_1}_F\left(\vec{q}^{\,2}_2\right)\,
\rightarrow\,F_{\rho NN}\left(\vec{q}^{\,2}_1\right)
\Delta^\rho_F\left(\vec{q}^{\,2}_1\right)\,
F_{a_1 NN}\left(\vec{q}^{\,2}_2\right)
\Delta^{\,a_1}_F\left(\vec{q}^{\,2}_2\right)\,,  \label{rhoa1ff}
\ee
in Eq.\,(\ref{Pbn}) in the case of the $\rho a_1 \gamma$ current. While in this case, the
static approximation is expected to be reasonably good, for the $\rho \pi \gamma$ current
the retardation current arises by the following change in the current $\vec{J}^{\,0}_\pi$,
\bea
\Delta^\rho_F\left(\vec{q}^{\,2}_1\right)\,\Delta^\pi_F\left(\vec{q}^{\,2}_2\right)\,
&\rightarrow&\,F_{\rho NN}\left(\vec{q}^{\,2}_1\right)
\Delta^\rho_F\left(\vec{q}^{\,2}_1\right)\,
F_{\pi NN}\left(\vec{q}^{\,2}_2\right)
\Delta^\pi_F\left(\vec{q}^{\,2}_2\right)\, \nonumber  \\
&&\,\times\left\{1\,+\,q^2_{10}\left[\Delta^\rho_F\left(\vec{q}^{\,2}_1\right)
-\frac{\partial}{\partial\vec{q}^{\,2}_1}\,ln\left(F_{\rho NN}
\left(\vec{q}^{\,2}_1\right)\right)\right] \right.  \nonumber \\
&&\left.+q^2_{20}\left[\Delta^\pi_F\left(\vec{q}^{\,2}_1\right)
-\frac{\partial}{\partial\vec{q}^{\,2}_1}\,ln\left(F_{\pi NN}
\left(\vec{q}^{\,2}_1\right)\right)\right]\right\}\,.  \label{jspiret} 
\eea
The retardation current $\vec{J}^{\,ret}_\pi$ is proportional to the 
energy transfers $q_{10}$ and $q_{20}$ defined as
\be
q_{i0}\,=\,\frac{1}{2M}\vec{q}_i \cdot \vec{P}_i\,,  \label{qi0}
\ee
which can be expressed via the c.m. variables by using Eqs.\,(\ref{cmc}). The part
of the current $\vec{J}^{\,ret}_\pi$ without the contribution from the derivative
of the form factors was derived earlier in \cite{Tal}.

The exchange charge densities are
\bea
J_{\,\pi\,0}\,&=&\,\frac{g_{\pi NN}}{2M}\,\frac{g}{2m_\rho}\,\frac{1+\kappa_V}{2 M}\,g_{\rho \pi \gamma}\,
K_{\rho \pi \gamma}(\vec{q}^{\,2})\,
[\vec{q}\cdot\vec{q}_2\times(\vec{\sigma}_1\times\vec{q}_1)]  \nonumber \\
& &  \hspace{35mm} \cdot\,(\vec{\tau}_1 
\cdot \vec{\tau}_2)\,\Delta^\rho_F(\vec{q}^{\,2}_1)\Delta^\pi_F(\vec{q}^{\,2}_2)\,+\,
\ot\,, \label{tjpi}  \\
J_{\,a_1\,0}\,&=&\,\frac{g_A}{2}\,g^2\,g_{\rho a_1 \gamma}\,\frac{1+\kappa_V}{2M}\left[K_1(\vec{q}^{\,2})
\vec{q}\,+\,K_2(\vec{q}^{\,2}\vec{q}_1\right]\cdot\left[\vec{\sigma}_2\times(
\vec{q}_1\times\vec{\sigma}_1)\right] \nonumber  \\
& &  \hspace{35mm} \cdot\,(\vec{\tau}_1 
\cdot \vec{\tau}_2)\,\Delta^\rho_F(\vec{q}^{\,2}_1)\,\Delta^{a_1}_F(\vec{q}^{\,2}_2)\,+\,
\ot\,.  \label{tja1}  
\eea

\section{Results and conclusions}
\label{reco}

In this paper, we have studied the electromagnetic processes in the $\pi \rho a_1 \omega$ system,
characterized by the change of the natural
parity of the participating particles, by using the anomalous chiral Lagrangian \cite{KM,STK}.
This Lagrangian contains 4 constants $\tilde{c}_7$--$\tilde{c}_{10}$, 
which were first extracted from the data in \cite{KM}. Using new data \cite{PDG} 
on the radiative decays of the $\rho$--, $\omega$-- and $f_1$ mesons
and on the decay $f_1\,\rightarrow\, \rho \pi \pi$ and the 
constant $\tilde{c}_7$ from \cite{KKW}, 
we have determined the constants $\tilde{c}_8$, $\tilde{c}_9$ and
$\tilde{c}_{10}$. 
The new constants differ considerably from the old ones. These constants enter the
decay rates in the combinations $\tilde{c}_7+\tilde{c}_9$ and $\tilde{c}_8+\tilde{c}_{10}$,
which can be obtained only up to the sign. We considered both possibilities and 
we extracted two sets of constants. 
Disregarding the sign problem, these constants are now known within a 10--15\% accuracy.
The main source of the uncertainty is due to an inconsistency in the data for the
radiative decays of the charged and neutral $\rho$ mesons and of the  $\omega$ meson.
However, the effective couplings g$_{\rho \pi \gamma}$ and g$_{\omega \pi \gamma}$ 
suffer from this inconsistency by an uncertainty only of about 4\%.

We have subsequently constructed the electromagnetic form factor 
$K_{\rho \pi \gamma}(q^2)$ that
contains, besides the well known term, 
an additional piece which can be large, if the corresponding choice of the constant $\tilde{c}_9$
is made. The presence of the $\rho a_1 \omega$ vertex in the Lagrangian gives rise to
the new $\rho a_1 \gamma$ vertex, which contains two form factors $K_1(q^2)$ and $K_2(q^2)$.

The isoscalar exchange currents, associated with these vertices, are constructed as well.
In addition to the known 
$\rho \pi \gamma$ current, we have also $\rho a_1 \gamma$ current with well
defined parameters extracted from the data. This is in contrast to $\omega \epsilon \gamma$
\cite{HT} or $\omega \sigma \gamma$ \cite{IG} currents based on less firm ground.
After the non--relativistic reduction,
relativistic corrections are also preserved in the space component of our currents,
in conformity with the results of Refs.\,\cite{HT},\cite{Tal}. 
As to the $\rho \pi \gamma$ current, the leading terms on the  right hand sides 
of Eqs.~(\ref{tjpi}) and (\ref{sjpi}) are the same as the exchange charge density (3.~2a) 
and the space component (3.2b) in
\cite{GH}, respectively. 
However, our electromagnetic form factor $K_{\rho\pi\gamma}(q^2)$ Eq.\,(\ref{rpgff})
exhibits an additional momentum dependence. 
Our relativistic corrections contain the contributions from 
the vertices, from the kinematical boost and from the retardation, with the part arising 
from the strong form factors included.
The current $\vec{J}_{a_1}$ Eq.\,(\ref{sja1}) and the density $J_{\,a_1\,0}$ Eq.\,
(\ref{tja1})are entirely new. 
The subsequent numerical work should reveal how much they can influence the deuteron form 
factors at high momentum transfers.

\section*{Acknowledgments}
This work is supported in part by the grant GA \v{C}R 202/00/1669. 
A part of this work was done during the stay of E.~T.~at the
Theoretical Physics Institute of the University of Alberta. He thanks Prof.~F.~C.~Khanna for
the warm hospitality. The research of F.~C.~K.~is supported in part by NSERCC.

We thank Prof. B. Mosconi for the correspondence and R. Teshima for the discussion.


\end{document}